# Theory of Rule 6 and it's Application to Round Robin Tournament

Pabitra Pal Choudhury, Sk. Sarif Hassan
Applied Statistics Unit,
Indian Statistical Institute, Kolkata, 700108, INDIA
Email: pabitrapalchoudhury@gmail.com
sarimif@gmail.com

Sudhakar Sahoo, Birendra Kumar Nayak
P.G. Department of Mathematics,
Utkal University, Bhubaneswar-751004, INDIA
Email: sudhakar.sahoo@gmail.com
bknatuu@yahoo.co.uk

*Abstract*— **In this paper we have used one 2 variable Boolean function called Rule 6 to define another beautiful transformation named as Extended Rule-6. Using this function we have explored the algebraic beauties and its application to an efficient Round Robin Tournament (RRT) routine for $2^k$ (k is any natural number) number of teams. At the end, we have thrown some light towards any number of teams of the form $n^k$ where n, k are natural numbers.**

## I. INTRODUCTION

In our earlier papers [1], [2] we have explored an application to self- similar and chaotic fractals formation, using one Boolean function named as 'Carry Value Transformation' (CVT) and in this paper, we have explored the algebraic beauties of some Boolean function and their application towards scheduling problem especially on *Round Robin Tournament* (RRT). We undertake the Boolean function of two variables, in particular named as Rule 6 [3]. Now interesting fact is that, for any two nonnegative integers we get a beautiful relation: $a + b = CVT(a,b) + \text{Extended Rule } 6(a,b)$. So in addition process, these two Boolean functions contribute the whole. More specifically, in the process of addition of two integers both the XOR (or Rule 6 in Wolfram's naming convention [3]) and XNOR (or Rule 9 in Wolfram naming convention [3]) operations are used and the relation is

$$a + b = 2.\text{Extended Rule} 9(a,b) + \text{Extended Rule} 6(a,b).$$

Moreover, these two functions have so many interesting properties and application towards real-world problems that are what we are about to explore in this paper. Earlier there were many research works in different domains and different viewpoints for characterizing the Boolean functions. However, we are using the Boolean functions in an unconventional paradigm. Clearly, there are $2^{2^2}=16$ Boolean functions of two variables and this Rule 6 is one of them which is of interest to us at the moment because it leads to an age old application like the efficient formation of *Round Robin Tournament* routine.

The organization of the paper is as follows. Section II discusses some of the fundamental concepts and earlier works such as Round Robin Tournament (RRT), finite group etc. which are used in the subsequent sections. The motivation of choosing those two Boolean functions, their algebraic properties and Extended Rule 6 definition are given in section III. Next, following our methodology it can be found in section IV the routine for 8 teams in RRT. Further investigation in section V leads us to get the RRT routine for $n^k$ (where n, k are natural numbers) number of teams. We next highlight future research directions and conclude the paper in section VI.

## II. REVIEW OF EARLIER WORKS AND FUNDAMENTAL CONCEPTS

Let us first warm up ourselves with some fundamentals, which are related to the current paper.

### A. Round Robin Tournament

A tournament of n different teams, where each team will play against the other exactly once, is said to be *Round Robin Tournament* [4].

The problem related to Round Robin Tournament is to make a routine. Suppose, the teams are labeled as $T_0, T_1, T_2...$

Now the team $T_i$ will play against $T_j$ in the $k^{th}$ round iff

$$i + j \equiv k \pmod{n}$$

It is to be noted that sometimes we need to make 'Bye' for some team, at some round due to the infeasibility of the game.

To make a routine for RRT we should keep in mind the followings…

The operational matrix has to be symmetric. Otherwise, tournament scheduling would be infeasible.

In each row or column of the operational matrix, each entry cannot be repeated. In other words, in each row or column each entry is unique.

To reduce the number of the round by 1, the main diagonal entries have to be same all through out the operational matrix.

Let us cite an example as follows…

Suppose there will be a Round Robin tournament for n= 8 teams. Let us make the routine by the method as just we have stated above.

Table 1: Shows the routine for 8 teams for Round Robin Tournament

| Team/Round | $T_0$ | $T_1$ | $T_2$ | $T_3$ | $T_4$ | $T_5$ | $T_6$ | $T_7$ |
|---|---|---|---|---|---|---|---|---|
| $R_0$ | B | $T_7$ | $T_6$ | $T_5$ | B | $T_3$ | $T_2$ | $T_1$ |
| $R_1$ | $T_1$ | $T_0$ | $T_7$ | $T_6$ | $T_5$ | $T_4$ | $T_3$ | $T_2$ |
| $R_2$ | $T_2$ | B | $T_0$ | $T_7$ | $T_6$ | B | $T_4$ | $T_3$ |
| $R_3$ | $T_3$ | $T_2$ | $T_1$ | $T_0$ | $T_7$ | $T_6$ | $T_5$ | $T_4$ |
| $R_4$ | $T_4$ | $T_3$ | B | $T_1$ | $T_0$ | $T_7$ | B | $T_5$ |
| $R_5$ | $T_5$ | $T_4$ | $T_3$ | $T_2$ | $T_1$ | $T_0$ | $T_7$ | $T_6$ |
| $R_6$ | $T_6$ | $T_5$ | $T_4$ | B | $T_2$ | $T_1$ | $T_0$ | B |
| $R_7$ | $T_7$ | $T_6$ | $T_5$ | $T_4$ | $T_3$ | $T_2$ | $T_1$ | $T_0$ |

Here it should be mentioned that we need exactly 8 rounds to complete Round Robin Tournament for 8 teams [4]. In addition, in each round one of the team get a 'Bye (B)'. We feel this 'Bye' is one type of loose end or inefficiency of the whole programme. Of course, sometime we need to make 'Bye' to the teams due to the infeasibility situation of the programme. However, if we could avoid unnecessary 'Byes' then it would be obviously efficient and excellent. Our next effort is to demonstrate an efficient routine for RRT.

### B. Boolean Functions and Wolframs Naming Convention

A Boolean function $f(x_1, x_2, ..., x_n)$ on $n$-variables is defined as a mapping from $\{0,1\}^n$ into $\{0,1\}$. It is also interpreted as the output column of its truth table $f$, which is a binary string of length $2^n$. For $n$-variables the number of Boolean functions is $2^{2^n}$ and each Boolean function is denoted as $f_R^n$ known as the function number $R$ (also interpreted as rule number $R$), in $n$-variable. Here $R$ is the decimal equivalent of the binary sequence (starting from bottom to top, with bottom is the LSB) of the function in the Truth Table, and numbering scheme is proposed by Wolfram and popularly known as Wolframs naming convention.

### C. Extended Rule 6

The Boolean function Rule 6, Rule 9 both are defined in the domain S= $\{0,1\}$ and range also S. Now we define another transformation named as Extended Rule 6, Extended Rule 9 in the extended set S= $\{0, 1, 2, ..., n : n=2^k$, for some natural number k$\}$.and range is also the extended set S. the details of the definitions and demonstration are given in the section 3.2.

### D. Analogous Rule 6

The Boolean function Rule 6 is defined in the domain S= $\{0,1\}$ and range is also S. Now we define another transformation named as Analogous Rule 6 in the set $\{0, 1, 2\}$ to itself. This definition we will consider as primal definition to define the extended Rule 6 in ternary logic system.

### E. Finite Group

A finite set along with a group structure, is known as Finite Group. Now for a given a finite set, it is not an easy task to say what would be the binary operation which could produce a group structure on it. So far, we know that under the usual binary operations like addition, multiplication, division etc the set of natural numbers ($N$) does not form a group. In this paper, we have defined one operation on any set of non-negative integers starting from 0, 1, 2, ... containing $2^k$ elements, $k \in N$ forming a group.

### III. TWO VARIABLE RULE-6 OR XOR OPERATOR

### A. Why We Choose Rule-6

Clearly, there are 16 rules of two variables. And there are some elegant algebraic properties associated with Rule 6, which are not observed in other rules. The rules are as follows…

Let us use four symbols

$$\alpha_0 = (0,0), \alpha_1 = (0,1), \alpha_2 = (1,0), \alpha_3 = (1,1)$$

Table 2: Shows 16 Boolean functions of two variables

| Rule no. / Nodes | 0 | 1 | 2 | 3 | 4 | 5 | **6** | 7 | 8 | 9 | 10 | 11 | 12 | 13 | 14 | 15 |
|---|---|---|---|---|---|---|---|---|---|---|---|---|---|---|---|---|
| $\alpha_0 = (0,0)$ | 0 | 1 | 0 | 1 | 0 | 1 | **0** | 1 | 0 | 1 | 0 | 1 | 0 | 1 | 0 | 1 |
| $\alpha_1 = (0,1)$ | 0 | 0 | 1 | 1 | 0 | 0 | **1** | 1 | 0 | 0 | 1 | 1 | 0 | 0 | 1 | 1 |
| $\alpha_2 = (1,0)$ | 0 | 0 | 0 | 0 | 1 | 1 | **1** | 1 | 0 | 0 | 0 | 0 | 1 | 1 | 1 | 1 |
| $\alpha_3 = (1,1)$ | 0 | 0 | 0 | 0 | 0 | 0 | **0** | 0 | 1 | 1 | 1 | 1 | 1 | 1 | 1 | 1 |

Clearly, there are 4 linear rules, which are Rule 0, Rule 6, Rule 10 and Rule 12 and the rest are non-linear rules. The notable matter is that the set $\{0, 1\}$ forms a group under the operation of Rule 6 and Rule 9. In this paper we have tried to define another operation on the extended domain S= $\{0, 1, 2 . . . n\}$ for any n of the form $2^k$ with the primal definition of Rule 6 and Rule 9 [see table 3a and 3b]. It should be noted that the set S does not forms a group under Extended Rule 9 operation, which we discuss in the next section. So in our extended scenario Rule 6 is the only best fitted rule. This group structure on any set $\{0, 1, 2 . . . n\}$ helps us to make the routine for Round Robin Tournament for any number of the form $2^n$ given teams. This is how we have been motivated on this algebraic property of Rule 6 and utilizing this rule, we have explored in the next section 4 an application to Round Robin Tournament. And in the next immediate subsection we discuss the algebraic properties of Rule 6 and Rule 9 in details.

*B. Algebraic properties of Rule 6*

The operation table for the set {0, 1} together with the operations on rule 6 and Rule 9 are as follows in the table 1 and table 2 respectively…

Table 3a: Primal definition of Rule 6

| Rule 6 | 0 | 1 |
|---|---|---|
| **0** | 0 | 1 |
| **1** | 1 | 0 |

Table 3b: Primal definition of Rule 9

| Rule 9 | 0 | 1 |
|---|---|---|
| **0** | 1 | 0 |
| **1** | 0 | 1 |

We observe the following facts from table-3 and table-4:

1. The set {0, 1} is closed under the rule 6, rule 9.
2. Associeativity holds good.
3. '0', '1' are the identity elements under rule 6 and rule 9 respectively.
4. Inverse of every element is itself for both the rules.
5. Commutativity holds good.

Altogether, we could say that the set {0, 1} forms a group under the rule 6 and Rule 9 operations. It could be seen that other 14 rules are unable to produce group structure on {0, 1}. In addition, it should be noted that rule 6 is a linear rule where as rule 9 is nonlinear. To compute the corresponding value of $(a,b)$ we need to express 'a' and 'b' in binary notations. Suppose that $a_{10} = (a_n, a_{n-1},...a_1)_2$ and $b_{10} = (b_n, b_{n-1},...b_1)_2$, where subscripts are denoting base of the numbers. We have known the value of

$\alpha_0 = (0,0)$, $\alpha_1 = (0,1)$, $\alpha_2 = (1,0)$, $\alpha_3 = (1,1)$

*Extended Rule* $6(a,b) = (a_n \oplus b_n, a_{n-1} \oplus b_{n-1},...,a_1 \oplus b_1)_2$
$= a \oplus b$, Where $\oplus$ denote the action of Rule 6 and $(a_n, b_n)$ is one of ($\alpha_0 = (0,0)$, $\alpha_1 = (0,1)$, $\alpha_2 = (1,0)$, $\alpha_3 = (1,1)$)

Let us demonstrate this with the help of an example. Let S be the set {0, 1, 2, 3, 4, 5, 6, 7}.

Let us compute the value of (5, 6) by Extended Rule 6.

Now, $5_{10} = (101)_2$ and $6_{10} = (110)_2$.

$5_{10} = (101)_2$

$6_{10} = (110)_2$

So, *Extended Rule* $6(5,6) = (1 \oplus 1, 0 \oplus 1, 1 \oplus 0) = (011)_2 = 3_{10}$

*1) Construction procedure of the operation table*

Arrange the elements of the set in ascending order starting from zero in both row and column wise.

Then compute the value corresponding to the each row and column entries by the following the above procedure and put the computed value to the corresponding position.

Therefore, the operational table is as follows…

Table 4: Shows the operational table of Extended Rule 6 on the set S

|   | **0** | **1** | **2** | **3** | **4** | **5** | **6** | **7** |
|---|---|---|---|---|---|---|---|---|
| **0** | 0 | 1 | 2 | 3 | 4 | 5 | 6 | 7 |
| **1** | 1 | 0 | 3 | 2 | 5 | 4 | 7 | 6 |
| **2** | 2 | 3 | 0 | 1 | 6 | 7 | 4 | 5 |
| **3** | 3 | 2 | 1 | 0 | 7 | 6 | 5 | 4 |
| **4** | 4 | 5 | 6 | 7 | 0 | 1 | 2 | 3 |
| **5** | 5 | 4 | 7 | 6 | 1 | 0 | 3 | 2 |
| **6** | 6 | 7 | 4 | 5 | 2 | 3 | 0 | 1 |
| **7** | 7 | 6 | 5 | 4 | 3 | 2 | 1 | 0 |

Now, from table-4, it is clear that

The set S is closed under the Extended Rule 6.

1. Associativity holds good.
2. Zero is the identity element.
3. Inverse of every element is itself.
4. Commutativity holds good.

Therefore, the set S= {0, 1, 2, 3, 4, 5, 6, 7} forms a group under the Extended Rule 6 as defined in the extended set S from {0, 1}. From the above table-4, it is obviously readable that the set {0, 1, 2, 3, 4... n} is closed under the defined Extended Rule 6 only if the cardinality of the set is of the form $2^k$ for k≥1. Another interesting fact of the table 4 is that the first and second quadrant is same as $3^{rd}$ and $4^{th}$ quadrant respectively of any square block of order $2^k \times 2^k$, k≥1. Therefore, from the knowledge of the initial definition of the Rule 6, consequently other higher square block could be organized very nicely. For example, see the upper left most $2^1 \times 2^1$ order block, it is copied to the $4^{th}$ quadrant in the $2^2 \times 2^2$ order block and $2^1$ is added in each entries of $2^1 \times 2^1$ order block and that transformed block is placed in the $2^{nd}$ and $3^{rd}$ quadrant of the $2^2 \times 2^2$ order block. In the similar way any higher order block could be organized inductively.

Now, to define the **Extended Rule-9**, we can make verbatim copy the definition just by replacing the word 'Rule-6' by 'Rule-9'. So let us compute the operation-table as we have done for Extended Rule-6.

Table 5: Shows the action of Extended Rule-9 on the set S

|   | 0 | 1 | 2 | 3 | 4 | 5 | 6 | 7 |
|---|---|---|---|---|---|---|---|---|
| **0** | 1 | 0 | 1 | 0 | 3 | 2 | 1 | 0 |
| **1** | 0 | 1 | 0 | 1 | 2 | 3 | 0 | 1 |
| **2** | 1 | 0 | 3 | 2 | 1 | 0 | 3 | 2 |
| **3** | 0 | 1 | 2 | 3 | 0 | 1 | 2 | 3 |
| **4** | 3 | 2 | 1 | 0 | 7 | 6 | 5 | 4 |
| **5** | 2 | 3 | 0 | 1 | 6 | 7 | 4 | 5 |
| **6** | 1 | 0 | 3 | 2 | 5 | 4 | 7 | 6 |
| **7** | 0 | 1 | 2 | 3 | 4 | 5 | 6 | 7 |

Clearly, this table would not be helpful for our purpose that is to make the routine for RRT as operation table or matrix is not satisfying the conditions as we have mentioned in section 2.1.1.

Initially, Rule 6 and Rule 9 both have enjoyed same structures on {0, 1}. However, in the extended domain the behavior of these two functions is different.

## IV. CONSTRUCTION OF ROUTINE OF ROUND ROBIN TOURNAMENT

Without loss of generality let us make the routine for the 8 ($2^3$ for k=3) teams for the Round Robin Tournament.

Let us use the information from Table 4 to find out the scheduling of the routine.

Let us define that the **round** for the $i^{th}$ team and $j^{th}$ team is determined by the action of Extended Rule 6 on (i, j).

Therefore, the routine is as follows…

Table 6: Routine for 8 teams due to Rule-6

| Team/Round | $T_0$ | $T_1$ | $T_2$ | $T_3$ | $T_4$ | $T_5$ | $T_6$ | $T_7$ |
|---|---|---|---|---|---|---|---|---|
| $R_1$ | $T_1$ | $T_0$ | $T_3$ | $T_2$ | $T_5$ | $T_4$ | $T_7$ | $T_6$ |
| $R_2$ | $T_2$ | $T_3$ | $T_0$ | $T_1$ | $T_6$ | $T_7$ | $T_4$ | $T_5$ |
| $R_3$ | $T_3$ | $T_2$ | $T_1$ | $T_0$ | $T_7$ | $T_6$ | $T_5$ | $T_4$ |
| $R_4$ | $T_4$ | $T_5$ | $T_6$ | $T_7$ | $T_0$ | $T_1$ | T-2 | $T_3$ |
| $R_5$ | $T_5$ | $T_4$ | $T_7$ | $T_6$ | $T_1$ | $T_0$ | $T_3$ | $T_2$ |
| $R_6$ | $T_6$ | $T_7$ | $T_4$ | $T_5$ | $T_2$ | $T_3$ | $T_0$ | $T_1$ |
| $R_7$ | $T_7$ | $T_6$ | $T_5$ | $T_4$ | $T_3$ | $T_2$ | $T_1$ | $T_0$ |

That is, there will be total seven rounds for the Round Robin tournament with 8 teams where no 'Bye' condition is attached. And there are four matches at each round of the tournament.

The allotted routine is given as follows…

Round 1:

1. Team 0 will play against the team 1 and vice versa.
2. Team 2 will play against the team 3 and vice versa.
3. Team 4 will play against the team 5 and vice versa.
4. Team 6 will play against the team 7 and vice versa.

Now, we are sufficiently warmed up the above routine for the other rounds for RRT. Therefore, we have reduced one round and without giving any 'Bye'.

Now one rational question may be raised that whether the group structure is necessary to obtain the RRT routine. The answer is not affirmative. Just keeping in mind the above stated necessary conditions (2.2.1) we can make one table, which can also produce such efficient Routine for RRT without having the group structure. For an example, let us give one such operational table:

If we see back the section 2.2 intuitively then it would be clear that our proposed method is suitable for the team of number $2^k$. But, it is not that for any given number it would not be possible to make the routine. However, ours is not an ideal routine in this sense. So in this paradigm, we need to discover some other technique to achieve an efficient RRT routine for arbitrary number of teams of the form $n^k$.

## V. ANALOGOUS-RULE 6 IN TERNARY LOGICAL DOMAIN

Let us assume the number of team is of the form $3^k$. Let us try to give a pathway towards this direction. Earlier we were using the two variable (0 and 1) Boolean functions. Now we need to use three-valued function. In this paradigm, we have crossed the boundary of Boolean algebra as we are considering three variable say as 0, 1 and 2.

Here,

$\alpha_0 = (0,0)$, $\alpha_1 = (0,1)$, $\alpha_2 = (0,2)$, $\alpha_3 = (1,0)$, $\alpha_4 = (1,1)$, $\alpha_5 = (1,2)$, $\alpha_6 = (2,0)$, $\alpha_7 = (2,1)$, $\alpha_8 = (0,0)$, $\alpha_9 = (2,2)$

The definition of the rule is as follows…

Table 7: Primal definition of analogous rule 6 in ternary domain

| Rule | 0 | 1 | 2 |
|---|---|---|---|
| **0** | 0 | 1 | 2 |
| **1** | 1 | 2 | 0 |
| **2** | 2 | 0 | 1 |

Clearly, the set {0, 1, 2} together with the binary operation due to *Analogous Rule 6* is a group.

Let us apply Analogous Rule 6 on the set S= {0, 1, 2, 3, 4, 5, 6, 7, 8} and we are obtaining the following table…

Table 8: Shows the action of Analogous Rule 6 on the set S

| **Rule** | **0** | **1** | **2** | **3** | **4** | **5** | **6** | **7** | **8** |
|---|---|---|---|---|---|---|---|---|---|
| **0** | 0 | 1 | 2 | 3 | 4 | 5 | 6 | 7 | 8 |
| **1** | 1 | 2 | 0 | 4 | 5 | 3 | 7 | 8 | 6 |
| **2** | 2 | 0 | 1 | 5 | 3 | 4 | 8 | 6 | 7 |
| **3** | 3 | 4 | 5 | 6 | 7 | 8 | 0 | 1 | 2 |
| **4** | 4 | 5 | 3 | 7 | 8 | 6 | 1 | 2 | 0 |
| **5** | 5 | 3 | 4 | 8 | 6 | 7 | 2 | 0 | 1 |
| **6** | 6 | 7 | 8 | 0 | 1 | 2 | 3 | 4 | 5 |
| **7** | 7 | 8 | 6 | 1 | 2 | 0 | 4 | 5 | 3 |
| **8** | 8 | 6 | 7 | 2 | 0 | 1 | 5 | 3 | 4 |

Clearly, {0, 1, 2, 3, 4, 5, 6, 7, 8} $3^2$ number of elements form a group under the defined operation.

Now let us make the routine for the tournament as follows…

Table 9: Routine for 8 teams due to Rule-6

| Team/Round | $T_0$ | $T_1$ | $T_2$ | $T_3$ | $T_4$ | $T_5$ | $T_6$ | $T_7$ | $T_8$ |
|---|---|---|---|---|---|---|---|---|---|
| $R_1$ | $T_1$ | $T_0$ | B | $T_7$ | $T_6$ | $T_8$ | $T_4$ | $T_3$ | $T_8$ |
| $R_2$ | $T_2$ | B | $T_0$ | $T_8$ | $T_7$ | $T_6$ | $T_5$ | $T_4$ | $T_3$ |
| $R_3$ | $T_3$ | $T_5$ | $T_4$ | $T_0$ | $T_2$ | $T_1$ | B | $T_8$ | $T_7$ |
| $R_4$ | $T_4$ | $T_3$ | $T_5$ | $T_1$ | $T_0$ | $T_2$ | $T_7$ | $T_6$ | B |
| $R_5$ | $T_5$ | $T_4$ | $T_3$ | $T_2$ | $T_1$ | $T_0$ | $T_8$ | B | $T_6$ |
| $R_6$ | $T_6$ | $T_8$ | $T_7$ | B | $T_5$ | $T_4$ | $T_0$ | $T_2$ | $T_1$ |
| $R_7$ | $T_7$ | $T_6$ | $T_8$ | $T_4$ | $T_3$ | B | $T_1$ | $T_0$ | $T_2$ |
| $R_8$ | $T_8$ | $T_7$ | $T_6$ | $T_5$ | B | $T_3$ | $T_2$ | $T_1$ | $T_0$ |

So fruitfully, we have made an efficient routine for the number of the team $3^k$ for k=2. Similarly for any number of the teams of the form any positive integral power of 2,3,… could be obtained by using the higher valued logic. It is to be noted that in table 9 we are observing 'Bye' only ones in each round but it has become efficient.

## VI. CONCLUSION AND FUTURE RESEARCH DIRECTIONS

We have used one Boolean function Rule 6 in the formation of a routine of RRT fruitfully. And we observed efficient RRT plan for number of teams of the form where n . But naturally there are so many numbers which could not be expressible in the form So, it would be our future effort to tackle this problem fruitfully for arbitrary number of teams. Authors are in firm conviction that our proposed method to construct operation table will help to generate SUDUKU problems and answers.